\documentclass[a4paper,UKenglish,cleveref, autoref, thm-restate]{oasics-v2021}

\hideOASIcs
\pdfoutput=1
\usepackage[utf8]{inputenc}
\usepackage{amsmath}
\usepackage{amsfonts}
\usepackage{mathtools}
\usepackage{tabularx}
\usepackage{booktabs}
\usepackage{xspace}
\usepackage{boxedminipage}
\usepackage[inline]{enumitem}
\usepackage{amsthm}
\usepackage{multirow}
\usepackage{pifont}
\usepackage{makecell}
\usepackage{adjustbox}
\usepackage{hyperref}
\usepackage[table]{xcolor}
\usepackage{tikz}
\usepackage{listings}

\newcommand{\sig}[1]{\langle #1 \rangle}
\newcommand{\node}[1]{\ensuremath{P_{#1}}\xspace}
\newcommand{\pipelined}{Pipelined Moonshot\xspace}
\newcommand{\Propose}{\mathsf{propose}}
\newcommand{\OptPropose}{{\sf opt\text{-}propose}}
\newcommand{\FallbackPropose}{{\sf fb\text{-}propose}}

\newcommand{\Timeout}{\mathsf{timeout}}
\newcommand{\Vote}{\mathsf{vote}}
\newcommand{\OptVote}{\mathsf{opt\text{-}vote}}
\newcommand{\FallbackVote}{\mathsf{fb\text{-}vote}}
\newcommand{\Cert}{\mathcal{C}}

\newcommand{\TimeoutMessage}{\mathcal{T}}
\newcommand{\TimeoutCert}{\mathcal{TC}}

\newcommand{\ViewTimer}{{\sf view\text{-}timer}}

\newcommand{\Lock}{{\sf lock}}

\newcommand{\TimeoutView}{{\sf timeout\_view}}

\title{Formally Verifying the Safety of Pipelined Moonshot Consensus Protocol}
\titlerunning{Formally Verifying the Safety of Pipelined Moonshot Consensus Protocol}
\author{M.~Praveen}{Chennai Mathematical Institute \and ReLaX}{praveenm@cmi.ac.in}{https://orcid.org/0000-0002-5734-7115}{Funded by Supra}
\author{Raghavendra Ramesh}{Supra Research}{r.ramesh@supraoracles.com}{https://orcid.org/0000-0002-6289-9723}{}
\author{Isaac Doidge}{Supra Research}{i.doidge@supraoracles.com}{https://orcid.org/0009-0008-3989-5901}{}
\authorrunning{M.~Praveen, Raghavendra Ramesh, Isaac Doidge}
\Copyright{Supra. This work, as part of the collaborative efforts of M.~Praveen, Raghavendra Ramesh and Isaac Doidge, falls under the intellectual property rights asigned to Supra in accordance with their agreements}
\acknowledgements{We acknowledge and thank Chandradeep Dey and Namrata Reddy, who were part of this project during its initial phase. We acknowledge Supra Research for funding M.~Praveen, the academic partner of this project and providing other support.}
\supplement{}
\supplementdetails[subcategory={Source-code}, cite={}, swhid={}]{Model}{https://github.com/Entropy-Foundation/suprabft-fv/tree/master/suprabft}
\nolinenumbers %uncomment to disable line numbering

\EventEditors{Bruno Bernardo and Diego Marmsoler}
\EventNoEds{2}
\EventLongTitle{5th International Workshop on Formal Methods for Blockchains (FMBC 2024)}
\EventShortTitle{FMBC 2024}
\EventAcronym{FMBC}
\EventYear{2024}
\EventDate{April 7, 2024}
\EventLocation{Luxembourg City, Luxembourg}
\EventLogo{}
\SeriesVolume{118}
\ArticleNo{8}
\begin{document}

	\maketitle
	\begin{abstract}
Decentralized Finance (DeFi) has emerged as a contemporary competitive as well as complementary to traditional centralized finance systems. As of 23rd January 2024, per Defillama~\cite{defillama} approximately USD 55 billion is the total value locked on the DeFi applications on all blockchains put together.

A Byzantine Fault Tolerant (BFT) State Machine Replication (SMR) protocol, popularly known as the consensus protocol, is the central component of a blockchain. If forks are possible in a consensus protocol, they can be misused to carry out double spending attacks and can be catastrophic given high volumes of finance that are transacted on blockchains. Formal verification of the safety of consensus protocols is the golden standard for guaranteeing that forks are not possible. However, it is considered complex and challenging to do. This is reflected by the fact that not many complex consensus protocols are formally verified except for Tendermint~\cite{tendermint-fv} and QBFT~\cite{qbft-fv}.

We focus on Supra's Pipelined Moonshot consensus protocol. Similar to Tendermint's formal verification, we too model Pipelined Moonshot using IVy and formally prove that for all network sizes, as long as the number of Byzantine validators is less than 1/3, the protocol does not allow forks, thus proving that Pipelined Moonshot is safe and double spending cannot be done using forks. The IVy model and proof of safety is available on~\cite{moonshot-fv}.
	\end{abstract}
\newcommand{\prcs}{\mathcal{V}}
\newcommand{\blkchn}{\mathbf{B}}

\section{Introduction}
Public blockchains are revolutionising modern society by rebranding traditional services mainly the traditional finance based services, and offering them on a ``decentralized trust'' platform. Here no single entity need be trusted as the network is typically open for permissionless participation and tolerates malicious behaviour of the participants up to a certain threshold. Though blockchains are being adopted by multiple domains of applications, finance or Decentralised Finance (DeFi), happens to be the {\em killer} application that has shot the blockchain technology to fame as well as towards a popular adoption. As of 23rd January 2024, per Defillama~\cite{defillama} approximately USD 55 billion is the total value locked on the DeFi applications on all blockchains put together.

Every node in a blockchain network runs a \emph{consensus protocol}, more precisely known as \emph{state machine replication (SMR)} protocol, that enables that node to transition from one blockchain state to the next in a consistent way so that no two nodes in the network end up in different states after processing the same sequence of transitions. The transitions are the clients' submitted ledger transactions that are batched into a block. The sequence of transitions form the chain of blocks, hence the name blockchain. We are interested in consensus protocols in a \emph{partially synchronous network} setting. In this setting it is well known that an SMR protocol tolerates up to one-third of the network nodes being Byzantine -- nodes that may crash or deviate arbitrarily from the protocol but are assumed to be unable to break cryptographic primitives like signatures. 

Many such protocols have been proposed as well as successfully been adopted in practice such as~\cite{tendermint, pbft, hotstuff, jolteon, ibft} but only a few protocols have been formally verified to the best of our knowledge: Both the safety and liveness of Tendermint~\cite{tendermint-fv} have been formally verified using Microsoft IVy~\cite{ivy}. The safety of QBFT (called IBFT earlier) has been verified~\cite{qbft-fv} using Microsoft's Dafny~\cite{dafny}.     

% Te New ones are being proposed, to improve certains aspects. Two such aspects are block period and commit latency. Block period is the delay between consecutive block proposals. Commit latency is the delay between proposal of a block and achieving $2/3$ majority of processes committing the block. Pipelined Moonshot protocol achieves better block period and commit latency compared to other similar protocols [citations].
Pipelined Moonshot~\cite{moonshot} is a novel rotating leader-based Byzantine fault tolerant SMR protocol that leverages {\em optimistic proposals} to achieve a high block throughput -- one block per network hop, and the lowest block finalization latency -- 3 network hops, in the scenario of a normal path. 
% In order to extract such a high performance in the normal path and to achieve the lowest {\em view length}, the fallback path becomes complex in terms of protocol logic.  
% On the other hand, distributed systems are inherently complex and trying to squeeze as little block period and commit latency as possible makes them even more complicated. 
It is well known that designing protocols and proving them correct by hand are notoriously prone to errors as many critical errors have been found even in peer-reviewed distributed protocols (see \cite{redbelly-fv} and references therein). In this paper, we focus on formally proving the safety of this protocol. 

{\em Safety} of a BFT SMR protocol is a critical requirement, ensuring that any two honest processes agree on the set of transactions executed and the order in which they are executed. Formally, if two honest processes have committed chains of blocks, then one of the chains must be (not necessarily strict) prefix of the other one. When a protocol loses safety, forks in the blockchain are possible, essentially yielding to the possibility of {\em double spending} which is catastrophic to the finances built on top of this blockchain. Pipelined Moonshot protocol is proved to be safe and live~\cite{moonshot} with a handwritten proof. The goal of this project is to provide a proof of safety in a formal verification tool.

\medskip

\noindent{\bf Our Contributions:}
\begin{itemize}
	\item We provide formal specification of the Pipelined Moonshot protocol in IVy~\cite{ivy}, serving as a reference for any implementation.
	\item We formally verify safety of Pipelined Moonshot successfully. This makes the formal specification a safety-error-free basis for any implementations to be developed.
	% \item Within IVy, we explicitly state and prove many facts that are used implicitly in the handwritten safety proof.
	% \item For proving properties in IVy, we need to manually write invariants of the protocol that are inductive and also imply the desired safety property that we wish to prove. We identify several properties the conjunction of which is proved to be inductive by IVy and also imply the desired safety property.
    \item We identify several invariants of the protocol to prove it safe. Invariants are useful in generating test cases to test implementations of the protocol \cite{zeng2009test, yuan2011test}.
    \item We record our experience in the form of challenges faced and the corresponding mitigations used. Learning from this experience we extend our wisdom as recommendations for applying formal verification to large projects.
\end{itemize}

%  CUTTING DOWN FOR SPACE
% The rest of the paper is structured as follows. Section~\ref{sec:rel} contrasts on the available methodologies for formal verification of consensus protocols and establishes our choice of IVy. Section~\ref{sec:setup} paints the high level overview of the Pipelined Moonshot protocol and details on the aim and scope of the formal verification. Section~\ref{sec:spec} explains the structure of the formal IVy specification of the Pipelined Moonshot protocol and the safety proof. Section~\ref{sec:cha} articulates the challenges in this endeavour and Section~\ref{sec:rec} shares the wisdom on applying the formal verification technique to other BFT SMR protocols in general. Finally Section~\ref{sec:conc} concludes this paper.

\section{Related Work}
\label{sec:rel}
In this section we detail various formal verification approaches for consensus protocol verification and motivate our choice of IVy for verifying the safety of Pipelined Moonshot protocol.

{\bf Model checking} approach typically models the protocol as some finite representation of a state transition system and expresses the correctness properties in some logic and enumerates exhaustively the state space validating against the given logical specifications. Various model checking tools like SPIN~\cite{spin}, TLC~\cite{tla}, Apalache~\cite{apalache} etc are popular. Typically for the consensus protocols of interest as long as the number of nodes in the system is not fixed the state space is unbounded and generally does not yield a decidable algorithm to model check. 
There are bounded model checking approaches where the number of nodes is fixed typically and that yields a finite state machine against which the correctness properties are checked. For instance, \cite{tendermint-fmbc} model checks the block synchronization protocol of Tendermint after fixing the number of nodes in the network using TLC and Apalache model checker. 
We are focused on the general problem of safety of the Pipelined Moonshot with no bounds on the network, hence model checking is not applicable and bounded model checking is not satisfactory. 

There are other approaches that identify a {\em small model property} in the given protocol verification problem and apply model checking against the small model. The small model property of a protocol $P$ essentially is a bound on the number of nodes, say $k$, such that the satisfaction of a property $\theta$ by $P$ when run with $k$ nodes imply that $P$ satisfies $\theta$ for all $n \ge k$. The threshold automata approach of~\cite{threshold} leverages this and builds a counter abstraction by counting the number of processes in each state. This has been applied to the verification of DBFT~\cite{redbelly-fv} asynchronous consensus protocol. However this approach is known to be hard and has not been applied so far for any of the  partially synchronous BFT consensus protocols, and we too could not find any direct ways of applying this approach to the safety verification of Pipelined Moonshot. 

We now turn to the deductive verification tools. In this approach, the protocol is modeled in some logic (typically first-order logic, its fragments or extensions) and the properties to be verified are also written in the same logic.  From these, formulas called Verification Conditions (VCs) are generated, whose unsatisfiability implies that the protocol has the desired property. With interactive theorem provers, proof of unsatisfiability is developed in a proof system (such as natural deduction system or its variants). With automated deductive verification tools, proof of unsatisfiability is given by Satisfiability Modulo Theory solver (SMT solver).

{\bf TLA+}~\cite{tla} supports a very expressive logic called TLA -- Temporal Logic of Actions, for specifying state machines and properties. 
% The \emph{blockchain synchronization} sub-protocol of the Tendermint protocol has been specified in TLA+ and verified~\cite{TendermintTLA2020} using TLAPS - TLA Proof System, an interactive theorem prover. 
We found that expressing the Pipelined Moonshot protocol in TLA+ to be very complex and huge, and so also the verification in TLAPS -- TLA Proof System, to be effortful as each and every lemma has to be proven more or less interactively. We were on the look out for solvers that push more automation and lessen the interaction with the solver. Another requirement of us was that the formal specification should be close to the real world programming languages so that the developer community may be comfortable using the formal specification as the basis for their implementation. We found the TLA+ specifications to be far from the interest of the developer community, unless they are trained specifically towards verification.

{\bf Dafny}~\cite{dafny} is a verification-aware programming language that facilitates specifying pre and post conditions for procedures and verify at compile time. Correctness by construction is the philosophy here. The code may also be compiled to regular programming languages like C\#, Java, JavaScript, Go and Python. The safety of QBFT (previously known as IBFT~\cite{ibft}) has been verified using Dafny.

It is also well known that the formal verification of distributed protocols is an arduous effort. Hence we were on the look out for a tool that maximally uses automation in proof building and we found {\bf IVy}~\cite{ivy} to fit the bill. IVy is a language and a tool for the formal specification and verification of distributed systems. IVy supports deductive verification using automated provers such as Z3~\cite{z3}, model checking, automated testing, manual theorem proving and generation of executable code. In order to achieve greater verification productivity, a key design goal for IVy is to allow the engineer to apply automated provers in the realm in which their performance is relatively predictable, stable and transparent. In particular IVy focuses on the use of decidable fragments of first-order logic. IVy supports modularisation of the specifications and proofs, aiding their readability and also ensuring that formulas passed to provers are in decidable fragments. This helps to some extent in getting the provers to return with answers quickly. 
% Formal verification is done by checking the (un)satisfiability of formulas using a Satisfiability Modulo Theory (SMT) solver like Z3~\cite{z3}. The syntax is closer to imperative programming languages usually used for implementing protocols. IVy emphasizes and drives the verification exercise towards  modularisation of the specifications and proofs, aiding their readability and also ensuring that formulas passed to the SMT solver are in decidable fragments. This helps to some extent in getting the SMT solver to return with an answer quickly.

As IVy embodies an imperative language, the protocol specification in IVy  serves as a sound reference for any implementation. Note that the safety of Tendermint has been verified using IVy~\cite{tendermint-fv}. For all these reasons we favoured IVy as the formal verification tool for verifying the safety of Pipelined Moonshot.

To make proofs easier, Pretend Synchrony~\cite{PretendSync} takes another route of reducing the problem of verifying asynchronous distributed protocols to the problem of verifying synchronous distributed protocols. However it has been applied only in the setting of crash faults setting but not in Byzantine faults setting, which is the focus of this paper.

\section{Safety of Pipelined Moonshot Consensus}
\label{sec:setup}
In this section we summarise \pipelined and elucidate the scope of our formal verification endeavour. 

\subsection*{\pipelined}
\pipelined~\cite{moonshot} is a chain-based, rotating leader Byzantine Fault Tolerant (BFT) State Machine Replication (SMR) protocol optimized for wide-area networks. It satisfies the safety and liveness properties of SMR under the partially synchronous network model~\cite{psync} given at most $f$ of the $n$ total participants in the protocol (which we will call \textit{validators}) are Byzantine, such that $f < \frac{n}{3}$. Without loss of generality, we assume that $n = 3f+1$ for the rest of the paper and use the term \emph{quorum} to refer to a set of $2f+1$ validators. The full details of \pipelined's setting are provided in~\cite{moonshot}.

% It generates a chain (linked-list) of blocks of client transactions (or some abstraction thereof) wherein each block explicitly references its immediate predecessor in the chain, called its \emph{parent}.

\begin{figure*}[!ht]
\small
    \begin{boxedminipage}[t]{1.08\textwidth}
    A \pipelined node $\node{i}$ runs the following protocol whilst in view $v$:
    \begin{enumerate}[leftmargin=*] 
        \item \textbf{Propose.} Upon entering $v$, the leader $L_v$ proposes using one of the following rules:
        \begin{enumerate}
            \item \textbf{Normal Propose.} If $L_{v}$ entered $v$ by receiving $\Cert_{v-1}(B_{k-1})$, multicast $\sig{\Propose, B_{k}, \Cert_{v-1}(B_{k-1}), v}$ such that $B_k$ extends $B_{k-1}$.

        \item \textbf{Fallback Propose.} If $L_{v}$ entered $v$ by receiving $\TimeoutCert_{v-1}$, multicast $\sig{\FallbackPropose, B_k, \Cert_{v'}(B_{k-1}), \TimeoutCert_{v-1}, v}$ such that $\Cert_{v'}(B_{k-1})$ is the highest ranked certificate in $\TimeoutCert_{v-1}$ and $B_k$ extends $B_{k-1}$. 
    \end{enumerate}

        \item \textbf{Vote.}
        $\node{i}$ votes at most twice in view $v$ when the following conditions are met:
        \begin{enumerate}
            \item \textbf{Optimistic Vote.} Upon receiving the first optimistic proposal $\sig{\OptPropose, B_{k}, v}$ where $B_k$ extends $B_{k-1}$, if (i) $\TimeoutView_i < v-1$, (ii) $\Lock_{i} = \Cert_{v-1}(B_{k-1})$ and (iii) $\node{i}$ has not voted in $v$, multicast an optimistic vote $\sig{\OptVote, H(B_k), v}_i$ for $B_k$.

            \item After executing \textit{Advance View} and \textit{Lock} with all embedded certificates, vote once when one of the following conditions are satisfied:
            \begin{enumerate}
                \item  \textbf{Normal Vote.} Upon receiving the first normal proposal $\sig{\Propose, B_{k}, \Cert_{v-1}(B_h), v}$, if (i) $\TimeoutView_i < v$, (ii) $B_k$ directly extends $B_h$ and (iii) $\node{i}$ has not sent an optimistic vote for an equivocating block $B'_{k'}$ in $v$, multicast $\sig{\Vote, H(B_k), v}_i$ for $B_k$.

                \item \textbf{Fallback Vote.} Upon receiving the first fallback proposal $\sig{\FallbackPropose, B_{k}, \Cert_{v'}(B_h), \TimeoutCert_{v-1},v}$ 
                if (i) $\TimeoutView_i < v$ and (ii) $B_k$ directly extends $B_h$ and $\Cert_{v'}(B_h)$ is the highest ranked certificate in $\TimeoutCert_{v-1}$, multicast $\sig{\FallbackVote, H(B_k), v}_i$ for $B_k$.
            \end{enumerate}
        \end{enumerate}

        \item \textbf{Optimistic Propose.} If $\node{i}$ is $L_{v+1}$ and voted for $B_{k}$ in view $v$, multicast  $\sig{\OptPropose, B_{k+1}, v+1}$ where $B_{k+1}$ extends $B_{k}$.

        \item \textbf{Timeout.} If $\ViewTimer_i$ expires and $\node{i}$ has not already sent $\TimeoutMessage_v$, then multicast $\sig{\Timeout, v, \Lock_i}_{i}$ and set $\TimeoutView_i = \max(\TimeoutView_i, v)$. Additionally, upon receiving $f+1$ distinct $\sig{\Timeout, v', \_}_*$ messages or $\TimeoutCert_{v'}$ such that $v' \ge v$ and not having sent $\TimeoutMessage_{v'}$, multicast $\sig{\Timeout, v', \Lock_i}_{i}$ and set $\TimeoutView_i = \max(\TimeoutView_i, v')$. 

        \item \textbf{Advance View.} $\node{i}$ enters $v'$ where $v' > v$ using one of the following rules:
        \begin{itemize}
            \item[-] Upon receiving $\Cert_{v'-1}(B_h)$. Also, multicast $\Cert_{v'-1}(B_h)$.
            \item[-] Upon receiving $\TimeoutCert_{v'-1}$. Also, unicast $\TimeoutCert_{v'-1}$ to $L_{v'}$.
        \end{itemize}
        Finally, reset $\ViewTimer_i$ to $3\Delta$ and start counting down.
    \end{enumerate}

    $\node{i}$ additionally performs the following actions in any view:
    \begin{enumerate}[leftmargin=*]
        \item \textbf{Lock.}\label{step:lock2} Upon receiving $\Cert_v(B_k)$ whilst having $\Lock_i = \Cert_{v'}(B_{k'})$ such that $v > v'$, set $\Lock_i$ to $\Cert_v(B_k)$.

        \item \textbf{Direct Commit.}\label{step:direct-commit-2} Upon receiving $\Cert_{v-1}(B_{k-1})$ and $\Cert_{v}(B_{k})$ such that $B_{k}$ extends $B_{k-1}$, commit $B_{k-1}$.
        
        \item \textbf{Indirect Commit.}\label{step:indirect-commit-2} Upon directly committing $B_{k-1}$, commit all of its uncommitted ancestors.
    \end{enumerate}
\end{boxedminipage}
\caption{The \pipelined Protocol~\cite{moonshot}}
\label{fig:pipelined-moonshot}
\end{figure*}

The protocol, presented in \Cref{fig:pipelined-moonshot} as given in~\cite{moonshot}, constructs a chain of blocks of client transactions (or some abstraction thereof) over a sequence of numbered views advanced by quorum decisions in the form of certificates. In \pipelined, a view, say $v$, may produce two types of certificates; a Quorum Certificate $\Cert_v(B_k)$ comprised of $2f+1$ votes for some block $B_k$ (where $k$ is the position or \emph{height} of $B$ in the blockchain) proposed for $v$, or a Timeout Certificate $\TimeoutCert_v$ comprised of $2f+1$ timeout messages for $v$. A \pipelined validator in view $v$ votes for $B_k$ when it receives $B_k$ in a valid proposal (described momentarily) and sends a timeout message for $v$, denoted $\TimeoutMessage_v$, that contains its locked QC---i.e., the QC with the highest view that it has observed so far---when it fails to exit $v$ before its view timer expires or when it observes evidence that at least one honest validator has already sent $\TimeoutMessage_v$. These rules together ensure that the protocol continually generates new certificates, preventing it from halting.

A validator that receives a certificate for view $v$ advances its local view to $v+1$ and resets its view timer. If it enters $v+1$ via a QC then it also locks the QC and multicasts it to ensure that its peers enter the view promptly. Otherwise, if it enters $v+1$ via $\TimeoutCert_v$ then it unicasts this certificate to the designated leader for $v+1$, denoted $L_{v+1}$, enabling it to enter the view and propose promptly. 

Upon entering $v+1$, $L_{v+1}$ creates a new block, say $B_l$, and multicasts it in a proposal that depends on the type of certificate it used to enter the view. If the view change was triggered by $\Cert_v(B_k)$, then $B_l$ \emph{directly extends} $B_k$ (i.e. $l = k+1$ and $B_l$ contains the hash digest of $B_k$) and $L_{v+1}$ multicasts a Normal Proposal containing both $B_l$ and $\Cert_v(B)$. Otherwise, $B_l$ extends the block certified by the QC with the highest view included in $\TimeoutCert_v$ and $L_{v+1}$ multicasts a Fallback Proposal containing both $B_l$ and $\TimeoutCert_v$. A validator in $v+1$ that receives a proposal of either type from $L_{v+1}$ that is constructed as previously described and has yet to either send a vote for an equivocating block (as described in \Cref{fig:pipelined-moonshot}) or a timeout message for $v+1$, multicasts a vote of the corresponding type for $B_l$ for $v+1$. Importantly, \pipelined ensures that votes cannot be aggregated into a QC unless they have the same type.

Upon voting for $B_l$, if the validator is $L_{v+2}$ then it also creates an Optimistic Proposal for $v+2$ containing a new block that extends $B_l$, presuming that $B_l$ will be certified. A validator that receives this proposal votes for it once in $v+2$ if it has not yet voted in $v+2$ and it entered the view via $\Cert_{v+1}(B_l)$ without having sent a timeout message for $v+1$. Optimistic Proposals, a distinguishing feature of Moonshot protocols, allow votes for the current view to be disseminated in parallel with a proposal for the next view when both leaders are honest. Comparatively, prior protocols require a leader to receive a certificate for the previous view before proposing, inherently sequentializing these actions.

A validator that observes the certification of a block and its immediate successor in the chain for adjacent views commits the block by permanently appending it to its local copy of the blockchain.

% \pipelined focuses on minimizing the latency between the proposal of a block and its permanent inclusion in the blockchain (i.e., its \emph{commit latency}), the latency between the proposals of distinct honest validators (i.e., its \emph{view change block period}) and the amount of time that a validator needs to wait view length. It achieves a block commit latency of $3\delta$ and a block period of $\delta$ when consecutive leaders are honest, and achieves a view length of $3\Delta$ in general yet being optimistically responsive~\cite{hotstuff}.

% there exists a time called the Global Stabilization Time (GST) after which message delivery takes
% at most $\Delta$ time. We use $\delta$ to denote the actual delivery time,
% which naturally satisfies $\delta \le \Delta$ after GST.

\subsection*{Safety}
An SMR protocol is \emph{safe} if it ensures that no two validators commit divergent blockchains. Let the local blockchain of validator $\node{i}$ be denoted by $\blkchn_i$. More formally, the safety property states 
% \textbf{TODO: [Isaac] Cite this? We use a different definition in the arxiv paper. This is still fine but may need a reference.} 
that for every run of the protocol, and for each pair of honest validators $(\node{i},\node{j}) \in \prcs \times \prcs$, either $\blkchn_i$ is a (not necessarily strict) prefix of $\blkchn_j$ or vice-versa.

\medskip

The \pipelined~\cite{moonshot} paper contains the handwritten proof of safety and liveness. As well established in the literature some errors may potentially be present in the handwritten proofs that could go overlooked. A recent example is the Chord~\cite{chord} protocol for distributed hash tables which, despite having more than $6000$ citations, was shown incorrect by Zave~\cite{zave} almost a decade after its publication. Since only formal verification can conclusively guarantee the absence of errors in a protocol, we aimed at developing mechanically verifiable proofs of the safety of \pipelined.

\section{Formal Specification and Verification using IVy}
\label{sec:spec}
In this section, we first present some of the preliminaries of IVy modeling, secondly we present an high level overview of the formal IVy specification of the Pipelined Moonshot consensus protocol, and then finally present the structure of our safety proof.

\subsection{IVy modeling setup}
IVy is a language and a tool for the formal specification and verification of distributed systems. Systems are represented as state transition machines. States are multi-sorted first-order structures, with relations and functions. Transitions specify how the state is mutated. Any update definable in first-order logic is supported. Update instructions can be given in sequence one after another, giving the syntax the flavor of a developer-friendly imperative programming language. Multiple update instructions can be grouped together into an \emph{action}, a keyword in IVy that is used to denote state transition specifications.

The system under consideration is typically split into multiple modules, with internal states of a module not allowed to be modified directly by other modules. One module can call actions of another module, passing parameters. Modules can reason about one another using {\em assume-guarantee} specifications, which are formulas specifying properties of the modules' states. Properties of the overall system has to be proved by writing \emph{inductive invariants}, which are properties satisfying two conditions --- initiation and inductiveness. Initiation means that the initial state of the system satisfies the invariant. Inductiveness means that if any of the actions are executed in any state that satisfies the invariant, the resulting state also satisfies the invariant.

Modules in IVy, apart from modularising the protocol specification and proofs, serves another deeper purpose. Multiple formulas used in the proof may together necessitate the use of logics that are undecidable. Modules in IVy allow proving different properties in isolation, ensuring that formulas supporting one invariant are invisible to other modules. This will allow users to control which formulas are passed to the underlying SMT solvers together, so that all calls to the SMT solver are within decidable fragments of first-order logic.

Only a high level abstract specification of the protocol is modeled and verified. Some implementation details are hence modeled with Boolean abstractions. For example, timers used in the protocol are replaced by Boolean propositions that indicate whether or not a timer has expired. In the IVy model, the Boolean proposition can switch value anytime non-deterministically to simulate a timer getting expired, instead of tracking the actual time elapsed since the last reset. This is a sound abstraction for proving safety.

Another abstraction we have adapted from the literature is handling quorums \cite{DecidableFragmentsIvy}. The protocol specification mandates that a validator needs to receive messages from a super majority of all validators (two-thirds of the entire set) in order to achieve a quorum. Verifying this detail would require having arithmetic in the formulas passed on to SMT solvers, potentially affecting the solver's performance. Instead, what is modeled is the \emph{quorum intersection property} \cite{DecidableFragmentsIvy} --- any two quorums have at least one common honest validator. It is this property of quorums that are mainly used in correctness proofs and is modeled in IVy as an axiom, avoiding the usage of arithmetic.

Validators receive messages from the network and verify their authenticity by checking digital signatures. It is assumed that Byzantine validators cannot break cryptographic primitives and hence they cannot forge signatures of honest validators. Checking digital signatures is not modeled in IVy --- the model assumes messages sent by honest validators are authentic. The model also disallows byzantine validators to send messages on behalf of other honest validator, though they can send any kind of message on behalf of themselves or other byzantine validators, even if such a message is not mandated to be sent by the protocol specification.

\subsection{Pipelined Moonshot Specification}
We have published our IVy specification and the formal proof of safety of Pipelined Moonshot online on GitHub~\cite{moonshot-ivy}. Following are the main modules in our IVy specification of Pipelined Moonshot:

% Each module is in one file. The IVy tool accepts files with the extension \emph{.ivy}. One file can be included in others, facilitating the usage of common library components and modular design. Following are the modules in our IVy model.

\textbf{Types}: 
This module contains the declarations of the data types used in the IVy specification. The types $\mathit{round\_t}$, $\mathit{height\_t}$ are declared to be instances of $\mathit{ubd\_seq}$,  a	small modification of $\mathit{unbounded\_sequence}$, which are finite but unbounded total linear orders. Round is the technical term used in our IVy model for view as used in the protocol specification \cite{moonshot}. The type $\mathit{process\_index\_t}$ is declared to be an instance of $\mathit{iterable}$, which allows a collection of validators to be iterated in a loop in IVy models. The above types are used conventionally while modeling protocols in IVy. Other types declared correspond to message types specified in the protocol specification: $\mathit{block\_t}$, $\mathit{quorum\_t}$, $\mathit{qc\_t}$, $\mathit{tc\_t}$. Common properties of these types are also written in this module, including the quorum intersection axiom. 
% This module is in the file types.ivy.

\textbf{Network}: This module models the network through which validators interact. It is almost same as the network model in Tendermint's IVy model \cite{tendermint-fv}, except for the kind of messages that can be sent. Here the kind of messages that can be sent are \emph{normal proposal, fallback proposal, optimistic proposal, normal prepare, fallback prepare, optimistic prepare, quorum certificate, timeout certificate, timeout} and \emph{weak timeout certificate}. A \emph{timeout certificate} is a collection of \emph{timeout} messages from a two thirds majority of validators, whereas a \emph{weak timeout certificate} is a collection \emph{timeout} messages from a number of validators at least one more than the number of Byzantine validators. The network model is that of any asynchronous one, where messages can be dropped or delivered multiple times and/or out of order. 
% This module is in the file network.ivy. 

\textbf{Moonshot}: The IVy specification of the Pipelined Moonshot is provided in this module. State variables of individual validators are declared and updates to the state variables are performed in response to specific events as specified in Figure \ref{fig:pipelined-moonshot}. The details of this module are provided in Appendix~\ref{app:moonshotSpec}.

\textbf{Quorum verification}: In implementation, the integrity of a quorum of messages received by a validator is verified by checking digital signatures accompanying the messages. Here, the integrity is checked by verifying that all honest members of a quorum have actually sent the corresponding messages. It is done in this module using the concept of \emph{monitors} provided by IVy --- they are additional updates to state variables that are performed whenever an action is performed by the protocol. This module contains monitors that record prepare and timeout messages sent by the validators. When a validator receives a quorum or timeout certificate, its integrity is checked by verifying from the records that all honest members of the quorum have actually sent the corresponding prepare or timeout message. This can be thought of as some kind of a central authority with a global view of all validators, who records all messages sent by the validators. Of course there is no such central authority in real implementation; it is only modeled here for the sake of proving safety.

% The protocol model in moonshot.ivy does not refer to the records maintained by monitors in this file. Only the properties to be proved in the next module refer to these records. This module is in the file quorum\textunderscore{}verification.ivy.

\textbf{Safety}: The safety module specifies the desired safety property in the form of an inductive invariant. Numerous supporting invariants are also included here, as detailed in the next sub-section. Following is a code snippet, that states the main safety property.
\begin{verbatim}
isolate full_safety = {
    relation blockchain_prefix(N1:process_index_t, N2:process_index_t)

    # the latest block committed to b_v by N1 is equal to or ancestor of the 
    latest block committed by N2. All blocks committed by N1 are also 
    committed by N2. Any block committed by N2 but not by N1 is a descendant 
    of the latest block committed by N1
    definition blockchain_prefix(N1,N2) = ...

    # this is the full safety property of the pipelined moonshot protocol: for
    any two honest processors N1,N2 the chain committed by N1 is a prefix of
    N2 or vice versa
    invariant forall N1,N2:process_index_t. is_good(N1) & is_good(N2) -> 
    (blockchain_prefix(N1,N2) | blockchain_prefix(N2,N1))
} with block_t, verify_quorum, certified_block_ancestor_m1, 
all_ancestors_committed, committed_blocks_ancestors, 
latest_committed_ancestors, commit_to_chain, commit_to_chain_m1
\end{verbatim}
The definition of the relation blockchain\textunderscore{}prefix above is not shown fully due to lack of space, but its intention is captured in the comment above. The \emph{with} clause above lists the names of other isolates, containing invariants supporting this one.

Table~\ref{tab:satistics} provides some statistics of these modules. Note that a typical line in safety.ivy is much longer than those in other files, since one whole invariant is written in one line of safety.ivy. The files also have extensive comments serving the purpose of readability and documentation. There are a total of 190 invariants and 23 monitors. A rough estimate of the ratio of sizes of program code vs.~proof is 1:3. Verifying the safety of Pipelined Moonshot took about 140 man hours after the protocol specification itself was stabilized. About 10\% of this was needed to model the protocol and the rest to complete the proofs.
\begin{table}[h]
    \centering
    \rowcolors{2}{white}{gray!25}
    \begin{tabular}{lp{8cm}r}
         Module & Contents & Lines \\
         \hline
         Types & Extended data types & 338 \\
         Network & Network model & 110 \\
         Moonshot & Pipelined Moonshot SMR protocol & 642 \\
         Quorum verification & Validating messages sent by quorum members& 165 \\
         Safety & Inductive invariants proving safety& 1309 \\
         & Total & 2564
    \end{tabular}
    \caption{Modules and their Lines of code}
    \label{tab:satistics}
\end{table}

\subsection{Structure of the Safety Proof}

Mechanically-checked proofs are developed interactively in a dialogue between a Verification engineer and the proof assistant -- IVy. The engineer gives the desired specification of the model and of the property, IVy attempts to prove that the model satisfies the property. It may prove, then all is well, it may fail showing a counterexample, or it may not come back for a reasonable amount of time. When satisfiability fails it shows logical errors in the protocol. When it takes an unreasonable amount of time, the engineer has to creatively craft some lemmas that aids the machine in its proof search. This is the standard iterative approach of building mechanised proof.

% We now describe the structure of our mechanically verified proof.

IVy could not prove the safety specification directly (as is typical). We had to write all the intermediate lemmas given in \cite{moonshot} and many more. We first outline the main steps in the handwritten safety proof.
\begin{enumerate}
    \item If an honest validator executes direct commit of a block $B$ as given in point 2 at the bottom of Figure \ref{fig:pipelined-moonshot}, then any subsequent block that achieves a quorum is a descendant of $B$. This is proved in \cite[Lemma 2, Lemma 3]{moonshot}.
    \item If an honest validator commits a block, it also commits all of its ancestors. This is implicit in \cite{moonshot}.
    \item For any two blocks committed by an honest validator, one is an ancestor of the other. This does not directly correspond to any result stated in \cite{moonshot}, but essential in our IVy proof.
    \item If $B_i$ (resp.~$B_j$) is the latest block committed by an honest validator $v_i$ (resp.~$v_j$), then $B_i$ is an ancestor of $B_j$ or vice-versa. This is a corollary of item 1 above.
    \item If there were two blocks that were divergent, one would be an ancestor of the other (by item 3 above) and both would be ancestors of the latest committed block (by item 4 above). Hence, both would be committed by all honest validators (by item 2 above), contradicting the hypothesis that they are divergent. This argument is essentially the proof of \cite[Theorem 3]{moonshot}.
\end{enumerate}
% Most such supporting invariants are also present in this module. 
% This module is in the file safety.ivy.

% The safety property to be proved for the protocol is written as an invariant in the file safety.ivy. IVy verifies that the property is an inductive invariant. 
% To prove that the safety property is an inductive invariant, additional supporting invariants may be needed. To illustrate this, consider a protocol with two integer variables $a,b$, initialized to $0,1$ respectively. There is one step in the protocol, which updates $a$ to the old value of $b$ and updates $b$ to the old value of $a+b$. We would like to prove that in any reachable state, $a \le b$. While this is true, it is not an inductive invariant --- consider the state with $a=-1$ and $b=1$ (which satisfies the property $a \le b$). But after one execution of the protocol, the state changes to $a=1$, $b=0$, which no longer satisfies the property $a \le b$. This is not an anomaly, because the state with $a=-1$ and $b=1$ can never be reached. If we add $a \ge 0 ~ \& ~ b \ge 0$ as a supporting invariant, $a \le b$ becomes inductive relative to the supporting invariant. Starting from any state that satisfies both $a \ge 0 ~ \& b ~ \ge 0$ and $a \le b$, if the protocol is executed for one step, the resulting state also satisfies both the properties. Thus, to prove that $a \le b$ is an inductive invariant, we need to add $a \ge 0 ~ \& ~ b\ge 0$ as a supporting invariant.

% Similarly to prove safety property of the pipelined Moonshot protocol, several other supporting properties need to be added. 
IVy verifies that properties are inductive invariants by generating formulas in Finite Almost Uninterpreted (FAU) fragment of first-order logic and passing them on to Z3 \cite{DecidableFragmentsIvy}. Trying to prove too many properties in one step often degrades the performance of the SMT solver. To overcome this, IVy allows to group together a small number of properties in an \emph{isolate}, specifying other isolates as supporting invariants. When verifying one isolate, other isolates that it depends on are assumed to be true. The dependencies can be checked later. The safety invariants in our model are structured into several isolates. The top level of this structure follows the structure of the handwritten proof that is summarized above, and is illustrated below. Here, (5, \lstinline|full_safety|) means that the point 5 above is proved in the isolate \lstinline|full_safety|, likewise for other nodes.  The arrow from (5, \lstinline|full_safety|) to (3, \lstinline|committed_blocks_ancestors|) means that the isolate \lstinline|full_safety| depends on other isolates: \lstinline|committed_blocks_ancestors| being one of them.

\medskip

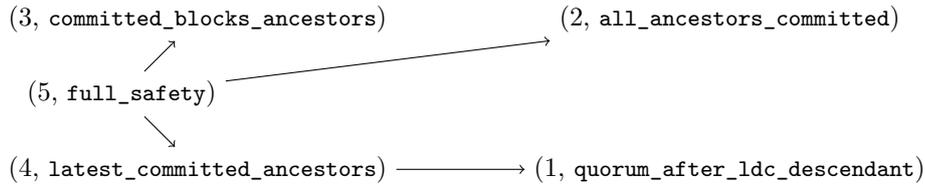
\begin{center}
    \begin{tikzpicture}
    \node (fs) at (0cm,0cm) {(5, \lstinline|full_safety|)};
    \node (lca) at ([xshift=1cm,yshift=-1cm]fs) {(4, \lstinline|latest_committed_ancestors|)};
    \node (cba) at ([xshift=1cm,yshift=1cm]fs) {(3, \lstinline|committed_blocks_ancestors|)};
    \node (aac) at ([xshift=7cm]cba) {(2, \lstinline|all_ancestors_committed|)};
     \node (qald) at ([xshift=7cm]lca) {(1, \lstinline|quorum_after_ldc_descendant|)};

     \draw[->] (fs) -- (cba);
     \draw[->] (fs) -- (aac);
     \draw[->] (fs) -- (lca);
     \draw[->] (lca) -- (qald);
    \end{tikzpicture}
\end{center}

The isolate \lstinline|quorum_after_ldc_descendant| is technically the most involved result in both the handwritten proof and IVy proof. This result is proved in IVy by induction on rounds. The principle of induction is taken to be an axiom and applied to the main invariant in the isolate \lstinline|quorum_after_ldc|. It states that if a block $B$ is committed directly by an honest validator, then any block proposed in later rounds that achieves a quorum has a parent proposed in the same round as $B$ or later rounds. Proving the invariants in the isolate \lstinline|quorum_after_ldc| itself is lengthy, indirectly involving around 30 other isolates.

% There are many other isolates that are not mentioned above. They consist of invariants that are used to support some of the above.

\section{Challenges}
\label{sec:cha}

The development and handwritten proof of safety and liveness of the pipelined Moonshot protocol underwent many cycles (some modifications to ensure liveness and some for simplifying the specification and proofs). This naturally resulted in iterating and refining the IVy specification too. The process of formally verifying safety (including analyzing counter examples given by IVy) uncovered some points in the specifications and proofs that were ambiguous and helped better understand many details that were implicit in the handwritten proofs. 

\medskip
\noindent Here we document some of the challenges faced in such a verification effort.

\paragraph*{Transitive closure} Ancestor relation is the binary transitive closure of the parent relation. For a validator to commit a block to its canonical chain, the block and its ancestors must have got quorums. The statement and proof of the property in the isolate quorum\textunderscore{}after\textunderscore{}ldc\textunderscore{}descendant uses the ancestor relation. Thus, many crucial parts of the model and safety proof depend on the ancestor relation. However, transitive closure of binary relations are not definable in first-order logic. To overcome this, we adapted a known technique \cite{DecidableFragmentsIvy}. If a binary relation is known to be the transitive closure of another base relation, then under some conditions the base relation can be defined from its transitive closure in first-order logic. To use this technique, a monitor in the isolate certified\textunderscore{}block\textunderscore{}ancestor\textunderscore{}m1 tracks when blocks get quorums and records the ancestor relation among them. Whenever a block gets quorum, the monitor updates the record, making the newly certified block a descendant of its parent block and all the parent block's ancestors. In the isolate certified\textunderscore{}block\textunderscore{}ancestor\textunderscore{}m5, we verify that the base relation obtained from the relation recorded by certified\textunderscore{}block\textunderscore{}ancestor\textunderscore{}m1 is indeed the parent relation. This challenge is not there for verifying Tenderming \cite{tendermint-fv}, where cross dependencies between isolates is lesser.

\paragraph*{Nested subroutine calls} As in most programming languages, actions in IVy can invoke other actions, which can themselves invoke more actions and so on. We observed empirically that with higher depth of nesting of action invocations, the performance of the IVy verifier slows down considerably. The protocol specification use subroutines that are called from multiple sites and the natural way to model it would be to have similar actions in IVy called from multiple actions. However, the slowdown in performance was significant enough that we resorted to inlining the subroutine calls. Further work is needed to understand the causes and more elegant workarounds.

\paragraph*{IVy verifier getting stuck without giving an answer} This challenge took up most of the time for executing this project. While verifying properties that were expected to be true, the IVy verifier would call the Z3 SMT solver, which would run for a long time without giving any answer. Such behaviour from SMT solvers cannot be entirely avoided, since they try to solve problems that have quite bad complexity theoretical lower bounds. There is no fixed template for handling this. Experience with using the tool and familiarity with the protocol being verified help a little bit. This is a challenge faced by most formal verification efforts; we felt it more since we had many invariants to prove, due to the complexity of the underlying protocol. Here are a few rules of thumb we resorted to, devised from trial and error.
\begin{description}
\item[Isolating the cause in the protocol] If the property being verified involved multiple steps in the protocol, we tried commenting out parts of the protocol and trying to verify the property. If the property was proved to be true/false after a particular section was commented out, then we could concentrate on that part to see what can be causing the SMT solver to diverge. This strategy helped us identify some subtle points that were implicitly assumed in the handwritten proof.
\item[Explicitly writing intermediate results] We illustrate this with an example. In the isolate quorum\textunderscore{}after\textunderscore{}ldc\textunderscore{}descendant\textunderscore{}m7, the third invariant states that under some conditions, the block $Bp$ is an ancestor of $B$. IVy could not prove this in a reasonable amount of time, so it was not clear whether it is due to lack of supporting invariants or because the SMT solver is diverging. We then added the first two invariants. The first one says that under the same condition, $Bp$ is an ancestor of $Bp1$ and the second one says that additionally, $Bp1$ is an ancestor of $B$. With the first two invariants added, IVy successfully verifies all the three in short time. 
% To prove that $Bp$ is an ancestor of $B$ (symbolically written as $B \rightarrow^* Bp$), we have added an intermediate block $Bp1$ and shown $B \rightarrow^*Bp1$ and $Bp1 \rightarrow^*B$. IVy then easily puts these two together to infer $B \rightarrow^*Bp$. The second invariant is bigger than the third one that we want to prove, but adding it actually helps speed up verification. 
There are many more examples like this. The main invariant of  quorum\textunderscore{}after\textunderscore{}ldc\textunderscore{}descendant\textunderscore{}m7 is inferred from a similar series of intermediate invariants, starting from quorum\textunderscore{}after\textunderscore{}ldc\textunderscore{}descendant\textunderscore{}m1, ending at quorum\textunderscore{}after\textunderscore{}ldc\textunderscore{}descendant\textunderscore{}m8 and then finally proving the invariant in quorum\textunderscore{}after\textunderscore{}ldc\textunderscore{}descendant.
\end{description} 

\section{Recommendations}
\label{sec:rec}
In this section we consolidate our experience with the safety verification of Pipelined Moonshot and attempt to distill some recommendations for applying formal verification techniques in proving the correctness of distributed systems.

\begin{itemize}
	\item Compared to semi-interactive theorem provers like Coq, the manual effort required with IVy is lesser. The proofs had to be flattened out into small steps manageable by SMT solvers, as explained in the last challenge. More research efforts like \cite{phaseStructures} are needed to reduce the burden of manually working out minute details, letting users concentrate on understanding protocols and correctness proofs intuitively.
	\item With complex protocols involving correctness proofs using hundreds of invariants, the success of deductive verifications tools that call SMT solvers in the background depends crucially on modularization of the proof, so that every SMT call is restricted to a small number of closely related formulas. For this, it is important at the outset to have a good idea of how the modules are going to be structured and which modules are meant for what. If this is lacking during the initial phase, chasing minute details during the verification process can quickly lead to huge, monolithic, incomprehensible and unmanageable pile of candidate invariants. In earlier attempts at verifying Pipelined Moonshot, we were sometimes in situations where we changed an invariant written earlier to suitably support a newly written invariant, only to realize later that this change affected an earlier dependency. We had lost track of which invariants supported which others and small changes in one invariant affected seemingly unrelated ones elsewhere.
	\item A related point is to be disciplined while establishing inter-dependencies among modules, specifically isolates in IVy. If  invariant 1 in isolate 1 needs invariant 2 in isolate 2 for support, it is tempting to mention the whole of isolate 2 as a dependency for isolate 1, instead of mentioning just invariant 2 of isolate 2. This may seem to be a time saver in the short term, but will result in unnecessary formulas (invariants in isolate 2 different from invariant 2) being passed to the SMT solver. Such unnecessary formulas can drastically degrade the performance of SMT solvers. Due to this, isolate 1 may pass all verification conditions in a short time currently but may not be able to do so in the future if additional invariants are added to isolate 2. If there is a group of invariants that are always together supporting other invariants, they should be recognized as such and grouped into an isolate, which is why this is related to the previous point of starting with well organized modules.
	\item The state of the art for formal verification of this scale very much requires experts with advanced knowledge of logic and related topics, who also need to understand the protocol being verified. An ideal team for formal verification would consist of experts with experience in using logic based verification tools on the one hand, and designers who understand the workings of the protocol at both abstract level and minute detail level on the other hand.
\end{itemize}

\section{Conclusion}
\label{sec:conc}
We have successfully verified the safety of a high performance and complex consensus protocol, namely Pipelined Moonshot, in IVy. This conclusively proves the absence of design or logic errors with respect to the protocol safety. Proving liveness is future work, possibly using \cite{LivenessToSafety2017} to reduce liveness to safety. 

This effort has yielded a developer friendly formal specification of the Pipelined Moonshot protocol that helps for any implementation of Pipelined Moonshot to safely base on.

We recorded our experience in the form of challenges faced and the mitigations employed during this project. Learning from this experience we enumerate some recommendations for applying formal verification for large distributed protocols. 

\bibliographystyle{plainurl}
\bibliography{refs}

\begin{thebibliography}{10}

\bibitem{moonshot-fv}
{IVy modeling of Pipelined Moonshot and its proof of safety}.
\newblock
  \url{https://github.com/Entropy-Foundation/suprabft-fv/tree/master/suprabft}.

\bibitem{spin}
{SPIN}.
\newblock \url{https://spinroot.com/spin/whatispin.html}.

\bibitem{tla}
{TLA+}.
\newblock \url{https://lamport.azurewebsites.net/tla/tla.html}.

\bibitem{tendermint-fv}
{Defillama}.
\newblock
  \url{https://galois.com/blog/2021/07/formally-verifying-the-tendermint-blockchain-protocol/},
  2021.

\bibitem{qbft-fv}
{Formal Verification of QBFT Safety}.
\newblock \url{https://github.com/Consensys/qbft-formal-spec-and-verification},
  2021.

\bibitem{defillama}
{Defillama}.
\newblock \url{https://defillama.com}, 2024.

\bibitem{moonshot-ivy}
{Moonshot Formal Verification in IVy - GitHub Repository}.
\newblock
  \url{https://github.com/Entropy-Foundation/suprabft-fv/tree/master/suprabft},
  2024.

\bibitem{z3}
{Z3 SMT Solver}.
\newblock \url{https://www.microsoft.com/en-us/research/project/z3-3/}, 2024.

\bibitem{tendermint-fmbc}
Sean Braithwaite, Ethan Buchman, Igor Konnov, Zarko Milosevic, Ilina
  Stoilkovska, Josef Widder, and Anca Zamfir.
\newblock {Formal Specification and Model Checking of the Tendermint Blockchain
  Synchronization Protocol}.
\newblock In Bruno Bernardo and Diego Marmsoler, editors, {\em 2nd Workshop on
  Formal Methods for Blockchains (FMBC 2020)}, volume~84 of {\em Open Access
  Series in Informatics (OASIcs)}, pages 10:1--10:8, Dagstuhl, Germany, 2020.
  Schloss Dagstuhl -- Leibniz-Zentrum f{\"u}r Informatik.
\newblock URL:
  \url{https://drops.dagstuhl.de/entities/document/10.4230/OASIcs.FMBC.2020.10},
  \href {https://doi.org/10.4230/OASIcs.FMBC.2020.10}
  {\path{doi:10.4230/OASIcs.FMBC.2020.10}}.

\bibitem{tendermint}
Ethan Buchman.
\newblock {\em Tendermint: Byzantine fault tolerance in the age of
  blockchains}.
\newblock PhD thesis, University of Guelph, 2016.

\bibitem{pbft}
Miguel Castro, Barbara Liskov, et~al.
\newblock Practical byzantine fault tolerance.
\newblock In {\em OSDI}, volume~99, pages 173--186, 1999.

\bibitem{moonshot}
Isaac Doidge, Raghavendra Ramesh, Nibesh Shrestha, and Joshua Tobkin.
\newblock Moonshot: Optimizing chain-based rotating leader bft via optimistic
  proposals, 2024.
\newblock \href {https://arxiv.org/abs/2401.01791} {\path{arXiv:2401.01791}}.

\bibitem{psync}
Cynthia Dwork, Nancy Lynch, and Larry Stockmeyer.
\newblock Consensus in the presence of partial synchrony.
\newblock {\em J. ACM}, 35(2):288–323, apr 1988.
\newblock \href {https://doi.org/10.1145/42282.42283}
  {\path{doi:10.1145/42282.42283}}.

\bibitem{phaseStructures}
Yotam~MY Feldman, James~R Wilcox, Sharon Shoham, and Mooly Sagiv.
\newblock Inferring inductive invariants from phase structures.
\newblock In {\em Computer Aided Verification: 31st International Conference,
  CAV 2019, New York City, NY, USA, July 15-18, 2019, Proceedings, Part II 31},
  pages 405--425. Springer, 2019.

\bibitem{jolteon}
Rati Gelashvili, Lefteris Kokoris-Kogias, Alberto Sonnino, Alexander
  Spiegelman, and Zhuolun Xiang.
\newblock Jolteon and ditto: Network-adaptive efficient consensus with
  asynchronous fallback.
\newblock In {\em FC}, pages 296--315, 2022.

\bibitem{apalache}
Igor Konnov, Jure Kukovec, and Thanh-Hai Tran.
\newblock Tla+ model checking made symbolic.
\newblock {\em Proc. ACM Program. Lang.}, 3(OOPSLA), oct 2019.
\newblock \href {https://doi.org/10.1145/3360549} {\path{doi:10.1145/3360549}}.

\bibitem{threshold}
Igor Konnov, Marijana Lazi\'{c}, Helmut Veith, and Josef Widder.
\newblock A short counterexample property for safety and liveness verification
  of fault-tolerant distributed algorithms.
\newblock In {\em Proceedings of the 44th ACM SIGPLAN Symposium on Principles
  of Programming Languages}, POPL '17, page 719–734, New York, NY, USA, 2017.
  Association for Computing Machinery.
\newblock \href {https://doi.org/10.1145/3009837.3009860}
  {\path{doi:10.1145/3009837.3009860}}.

\bibitem{dafny}
K.~Rustan~M. Leino.
\newblock Dafny: An automatic program verifier for functional correctness.
\newblock In Edmund~M. Clarke and Andrei Voronkov, editors, {\em Logic for
  Programming, Artificial Intelligence, and Reasoning}, pages 348--370, Berlin,
  Heidelberg, 2010. Springer Berlin Heidelberg.

\bibitem{DecidableFragmentsIvy}
Kenneth~L. McMillan and Oded Padon.
\newblock Deductive verification in decidable fragments with ivy.
\newblock In Andreas Podelski, editor, {\em Static Analysis}, pages 43--55,
  Cham, 2018. Springer International Publishing.

\bibitem{ibft}
Henrique Moniz.
\newblock The istanbul bft consensus algorithm, 2020.
\newblock \href {https://arxiv.org/abs/2002.03613} {\path{arXiv:2002.03613}}.

\bibitem{LivenessToSafety2017}
Oded Padon, Jochen Hoenicke, Giuliano Losa, Andreas Podelski, Mooly Sagiv, and
  Sharon Shoham.
\newblock Reducing liveness to safety in first-order logic.
\newblock 2(POPL), 2017.
\newblock \href {https://doi.org/10.1145/3158114} {\path{doi:10.1145/3158114}}.

\bibitem{ivy}
Oded Padon, Kenneth~L. McMillan, Aurojit Panda, Mooly Sagiv, and Sharon Shoham.
\newblock Ivy: safety verification by interactive generalization.
\newblock {\em SIGPLAN Not.}, 51(6):614–630, jun 2016.
\newblock \href {https://doi.org/10.1145/2980983.2908118}
  {\path{doi:10.1145/2980983.2908118}}.

\bibitem{MoonshotWhitePaper}
Supra Research.
\newblock Moonshot: Optimistic proposal for blockchain-based state machine
  replication.
\newblock \url{https://supraoracles.com/news/moonshot-consensus/}.

\bibitem{chord}
Ion Stoica, Robert Morris, David Karger, M.~Frans Kaashoek, and Hari
  Balakrishnan.
\newblock Chord: A scalable peer-to-peer lookup service for internet
  applications.
\newblock In {\em Proceedings of the 2001 Conference on Applications,
  Technologies, Architectures, and Protocols for Computer Communications},
  SIGCOMM '01, page 149–160, New York, NY, USA, 2001. Association for
  Computing Machinery.
\newblock \href {https://doi.org/10.1145/383059.383071}
  {\path{doi:10.1145/383059.383071}}.

\bibitem{redbelly-fv}
Pierre Tholoniat and Vincent Gramoli.
\newblock {\em Formal Verification of Blockchain Byzantine Fault Tolerance},
  pages 389--412.
\newblock Springer International Publishing, Cham, 2022.
\newblock \href {https://doi.org/10.1007/978-3-031-07535-3_12}
  {\path{doi:10.1007/978-3-031-07535-3_12}}.

\bibitem{PretendSync}
Klaus v.~Gleissenthall, Rami~G\"{o}khan K\i{}c\i{}, Alexander Bakst, Deian
  Stefan, and Ranjit Jhala.
\newblock Pretend synchrony: synchronous verification of asynchronous
  distributed programs.
\newblock {\em Proc. ACM Program. Lang.}, 3(POPL), jan 2019.
\newblock \href {https://doi.org/10.1145/3290372} {\path{doi:10.1145/3290372}}.

\bibitem{hotstuff}
Maofan Yin, Dahlia Malkhi, Michael~K Reiter, Guy~Golan Gueta, and Ittai
  Abraham.
\newblock Hotstuff: Bft consensus with linearity and responsiveness.
\newblock In {\em PODC}, pages 347--356, 2019.

\bibitem{yuan2011test}
Yuan Yuan, Zeng Fanping, Zhu Guanmiao, Deng Chaoqiang, and Xiong Neng.
\newblock Test case generation based on program invariant and adaptive random
  algorithm.
\newblock In {\em Advances in Information Technology and Education:
  International Conference, CSE 2011, Qingdao, China, July 9-10, 2011,
  Proceedings, Part I}, pages 274--282. Springer, 2011.

\bibitem{zave}
Pamela Zave.
\newblock Using lightweight modeling to understand chord.
\newblock {\em SIGCOMM Comput. Commun. Rev.}, 42(2):49–57, mar 2012.
\newblock \href {https://doi.org/10.1145/2185376.2185383}
  {\path{doi:10.1145/2185376.2185383}}.

\bibitem{zeng2009test}
Fanping Zeng, Qing Cao, Liangliang Mao, and Zhide Chen.
\newblock Test case generation based on invariant extraction.
\newblock In {\em 2009 5th International Conference on Wireless Communications,
  Networking and Mobile Computing}, pages 1--4. IEEE, 2009.

\end{thebibliography}
\newpage
\appendix
\section{IVy Specification of the Pipelined Moonshot Protocol}
\label{app:moonshotSpec}
The IVy specification of the Pipelined Moonshot is provided in the module \textbf{Moonshot}. The white paper \cite{MoonshotWhitePaper} by Supra research describes the same protocol as \cite{moonshot} but in a format that is better suited to serve as a starting point for implementation. The structure of our IVy model closely follows the description in \cite{MoonshotWhitePaper}, so we use it in the following to explain the orgnization of the IVy model.

The Moonshot module consists of declarations of state variables to be maintained by honest validators as given in \cite[Table II]{MoonshotWhitePaper}. These are then followed by \emph{actions}, the keyword in IVy used to denote updates to the state variables performed in response to specific events. Below we list the actions and the corresponding events specified in \cite{MoonshotWhitePaper}. Here, $f$ is the maximum number of Byzantine validators tolerated.
\begin{table}[h]
    \centering
    \rowcolors{2}{white}{gray!25}
    \begin{tabular}{llp{4.5cm}}
         Action & Subroutine in \cite{MoonshotWhitePaper} & Triggering event \\
         \hline
         \lstinline!qc_processing! & Algorithm 2 line 27 & Receiving a quorum certificate\\
         \lstinline!optimistic_proposal_processing! & Algorithm 2 line 37 & Receiving an\newline optimistic proposal\\
         \lstinline!normal_proposal_processing! & Algorithm 2 line 49 & Receiving a normal proposal\\
         \lstinline!timer_expire! & Algorithm 3 line 73 & Timer expires\\
         \lstinline!timeout_sync! & Algorithm 3 line 76 & Receiving timeout messages \newline from $f+1$ validators\\
         \lstinline!tc_processing! & Algorithm 3 line 80 & Receiving a timeout certificate\\
         \lstinline!fallback_proposal_processing! & Algorithm 3 line 89 & Receiving a fallback proposal
    \end{tabular}
    \caption{IVy actions and their corresponding subroutine in the specification}
    \label{tab:specs}
\end{table}

% The definition of when a block is local direct commit as specified in \cite[Definition 6]{MoonshotWhitePaper} is also written here. 
Following is a code snippet from the action \lstinline!normal_proposal_processing!, broadcasting a prepare message.
\begin{verbatim}
    # This function encodes the conditions necessary 
    for processing a normal proposal
    function send_prepare_n_condition(B_pr:block_t, QC:qc_t) : bool
    definition send_prepare_n_condition(B_pr, QC) = block_t.round(B_pr,r_c)
    & a_f < r_c & t_l < r_c & (a_o < r_c  | b_o = B_pr) &
    (forall B:block_t. forall R:round_t. qc_t.block(QC,B) & 
    block_t.round(B,R) -> block_t.parent(B_pr,B) & round_t.succ(R, r_c))
    
    # The procedure in Line 49 of Algorithm 2, executed upon receiving 
    # a normal proposal
    action normal_proposal_processing(b_pr:block_t, qc:qc_t) = {

        require received_proposal_n(b_pr, leader(r_c));
        require received_qc(qc);
        require block_t.cstd(b_pr);

        # Require that the timer has not yet expired for this round
        require ~ t_r;
        require ~ possessed_normal_for_round(r_c);

        #require that the parent of the proposed block b_pr is certified by
        # the accompanying QC qc
        require forall B:block_t. qc_t.block(qc,B) -> 
        block_t.parent(b_pr,B);
        
        possessed_normal_for_round(r_c) := true;

        # If the accompanying qc is not yet processed yet, do that first
        if some b:block_t. qc_t.block(qc,b) & ~ processed_qc(b) {
            call qc_processing(qc);
        }

        # After processing the accompanying the qc, require that the
        # conditions in lines 50-56 of Algorithm are met
        require send_prepare_n_condition(b_pr,qc);

        # This condition verified by IVy ensures that the parent of the
        # proposed block b_pr is for a strictly lesser round
        ensure block_t.parent(b_pr,Bp) & block_t.round(Bp,Rp) ->
        Rp < r_c;

        #Line 58: propose optimistic
        ###### proposeOptimistic ############
        # Line 7,8 of Algorithm 1
        var rs := round_t.next(r_c);
        if leader(rs) = id & b_o ~= b_pr{

            #Lines 10-11 of Alorithm 1
            var b := block_t.block(rs,b_pr);
            var m : msg;
            m.kind := msg_kind.proposal_o;
            m.block := b;
            m.src := id;

            call shim.broadcast(id, m);
        }

        if send_prepare_n_condition(b_pr,qc) {

            # Line 59 of Algorithm 2: broadcast prepare normal message
            var m : msg;
            m.kind := msg_kind.prepare_n;
            m.block := b_pr;
            m.src := id;

            call shim.broadcast(id, m);

            # Line 60 of Algorithm 2: a_n is updated to the current
            # round
            a_n := r_c;
        }
    }
\end{verbatim}
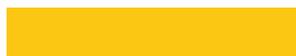
\end{document}